\documentclass[A4]{PoS}

\usepackage{subcaption}

\newcommand{\mnv}{MINER$\nu$A}

\let\OLDthebibliography\thebibliography
\renewcommand\thebibliography[1]{
  \OLDthebibliography{#1}
  \setlength{\parskip}{0pt}
  \setlength{\itemsep}{0pt plus 0.3ex}
}

\title{Implications of recent \mnv{} results for neutrino energy reconstruction}

\ShortTitle{Implications of recent \mnv{} results for neutrino energy reconstruction}

\author{\speaker{Jeremy Wolcott}, for the MINERvA collaboration \\
        University Of Rochester\\
        E-mail: \email{jwolcott@pas.rochester.edu}}

\abstract{Among the most important tasks of neutrino oscillation experiments is correctly estimating the parent neutrino energy from the by-products of their interactions.  Large uncertainties in our current understanding of such processes can significantly hamper this effort.  We explore several recent measurements made using the \mnv{} detector in the few-GeV NuMI muon neutrino beam at Fermilab: the differential cross-section vs. $Q^2$ for charged-current quasi-elastic scattering, the differential cross-sections vs. pion angle and pion kinetic energy for resonant single charged pion production, and the differential cross-sections vs. pion angle and kinetic energy for coherent pion production.  We furthermore discuss their implications for energy reconstruction in oscillation measurements.}

\FullConference{16th International Workshop on Neutrino Factories and Future Neutrino Beam Facilities\\
		 25 -30 August, 2014\\
		 University of Glasgow, United Kingdom}

\begin{document}

	\section{Introduction}
		The study of neutrino oscillation in accelerator-based experiments typically reduces to the reconstruction of the energy of the incident neutrino which induced a reaction inside a detector, $E_{\nu}$.  Unfortunately, no single reaction channel completely dominates in the energy range of interest (roughly $0.5 \le E_{\nu} \le 10$, in GeV).  Therefore, to successfully reconstruct the $E_{\nu}$ spectrum, it is usually necessary to employ a combination of detailed event selection criteria and a carefully chosen energy estimator which reproduces the neutrino energy for a given class of interactions.  But despite experimenters' best efforts, the complex and overlapping nature of the reaction mechanisms usually renders it impossible to avoid significant contamination from background processes.  Thus, it is essential to have accurate predictions for both the rate and spectral shape of not only the intended signal process, but all probable background reactions as well.
		
		Historically, the available models have been constructed from admittedly crude pictures of the heavy nuclei from which modern detectors are constructed.  Currently available simulation tools have performed adequately when compared to data for simple targets, such as deuterium\cite{ANL}; but studies on heavier materials such as hydrocarbons\cite{MiniBooNE CCQE} have yielded results that are murky at best.  As a result, model uncertainty has slowly crept to the head of lists of systematic uncertainties for these results.  Dedicated studies of the various neutrino reactions at neutrino energies of $\sim 1-10$ GeV have thus become an indispensible part of the neutrino oscillation program.
		
		The \mnv{} experiment, which uses the intense NuMI muon neutrino beam at Fermilab, is dedicated to measuring neutrino cross-sections as a function of various kinematic parameters on a range of nuclear targets.  This information can be compared directly to models and used to improve or select between them.  The detector itself consists of a central sampling scintillator region, built from strips of fluoror-doped scintillator glued into sheets, then stacked transverse to the beam axis; both barrel-style and downstream longitudinal electromagnetic and hadronic sampling calorimeters; and a collection of upstream passive targets of lead, iron, graphite, water, and liquid helium.  Sitting as it does just upstream of the MINOS near detector in NuMI, \mnv{} is also able to make use of the former as a muon spectrometer for most muons which exit the back of \mnv{}.  Much more detail on the detector design and performance is available elsewhere.\cite{MINERvA NIM}
	
	\section{Initial-state models: CCQE}
		Both of the event generators in widespread use in by modern neutrino experiments, GENIE\cite{GENIE} and NEUT\cite{NEUT}, begin from a model which pictures the nucleons within a nucleus as non-interacting fermions in a relativistic gas (the Relativistic Fermi Gas model, or RFG), and select \textit{post facto} from the plethora of available attempts at modeling interactions between them.  The extent to which these approximations are valid can be examined directly by making measurements of the simplest charged interaction neutrinos experience, that of the charged-current quasi-elastic (CCQE) process, and comparing the data to predictions from the various models.	In CCQE scattering of muon neutrinos, a W boson is exchanged between the neutrino and a neutron, resulting in a negatively charged muon and a proton in the final state: $\nu_{\mu} n \rightarrow \mu^{-} p$.  (The analogous process for muon antineutrino scattering reverses the lepton number and isospin: $\bar{\nu}_{\mu} p \rightarrow \mu^{+} n$.)  We analyze this reaction by selecting candidate events in two different ways.
	
		The first analysis\cite{1trk ccqe antinu paper,1trk ccqe nu paper} requires an event with a reconstructed track exiting \mnv{} and matching to a track in the MINOS near detector: this is the muon candidate.  (We restrict the sample to events which are analyzed as having the correct charge appropriate to neutrinos of the helicity selected by the NuMI beam configuration; samples were collected in both neutrino- and antineutrino-enhanced beams.)   We furthermore insist that there be no more than two isolated showers away from the muon vertex in the event for neutrino mode (only a single shower for antineutrinos).  With the muon energy and angle thus obtained, we are able to apply the following well-known estimators for the neutrino energy $E_{\nu}$ and square of the four-momentum transferred to the nucleus $Q^{2}$, assuming that an event was a CCQE interaction on a stationary nucleon (with $m_{\mu,n,p}$ the various masses, and $E_{b}$ a semi-empirical binding energy which accounts for the nuclear potential):
		\begin{equation}
			E_{\nu}^{QE} = \frac{m_{n}^{2} - (m_{p} - E_{b})^{2} - m_{\mu}^{2} + 2(m_{p}-E_{b}E_{\mu})}{2(m_{p} - E_{b} - E_{\mu} + p_{\mu} \cos{\theta_{\mu}})}
		\end{equation}
		\begin{equation}
			Q^{2}_{QE} = 2 E_{\nu}^{QE} \left(E_{\mu} - p_{\mu} \cos{\theta_{\mu}}\right) - m_{\mu}^{2}
			\label{eq:q2}
		\end{equation}
		
\begin{figure}[h]
			\centering
			\includegraphics[width=0.7\textwidth]{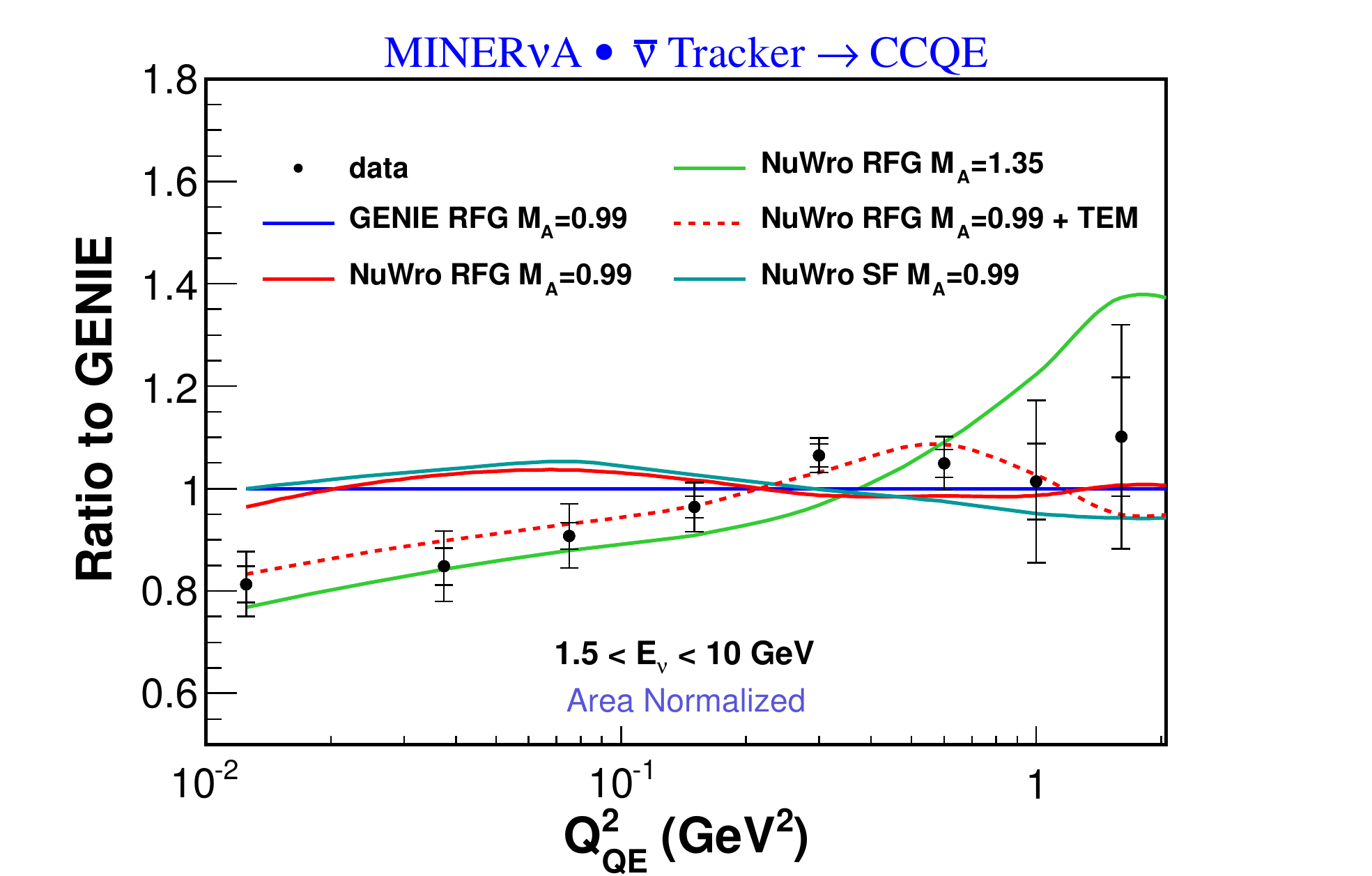}
			\caption{Ratios of the differential cross-section vs. $Q^{2}_{\mu}$ for \mnv{} data and several models to the usual RFG model as implemented in GENIE.  Other models considered: an alternative RFG implementation from the NuWro generator with different values of the free parameter $M_{A}$, a spectral function (SF) model, and the RFG coupled with transverse enhancement (TEM; described in the text).  The figure shows antineutrino scattering; neutrino scattering is similar.}
			\label{fig:1trk ccqe Q2}
		\end{figure}

		Our last selection requirement is an upper limit on the total non-muon (``recoil'') energy outside of a region immediately surrounding the muon vertex (for more on which see further below), which we make as a monotonically increasing function of the reconstructed $Q^{2}$.  Once reconstructed, we normalize the predicted backgrounds to the data by fitting them in a sideband region in the recoil energy.  Then, we subtract the predicted background counts from the data, un-smear the muon energy and angle using an iterative Bayesian unfolding technique.  Finally, we can normalize by the predicted efficiency and flux, and the number of neutrons (protons for anti-neutrinos) in our fiducial volume region to obtain a cross-section.  Eq. \ref{eq:dsigma} gives this in bins $i$ of $Q^{2}$, using $\epsilon$ for efficiency, $\Phi$ for the flux (integrated across the energy range of interest), $T_{n}$ for the number of targets, $\Delta_{i}$ for the bin width, and $U_{ij}$ for the predicted migration from bin $i$ into bin $j$:
		\begin{equation}
			\label{eq:dsigma}
			\left( \frac{d\sigma}{dQ^{2}_{QE}} \right)_{i} = \frac{1}{\epsilon_{i} \Phi T_{n} \Delta_{i}} \times \sum_{j}{U_{ij} \left(N_{j}^{\mathrm{data}} - N_{j}^{\mathrm{bknd\ pred}}\right)}
		\end{equation}
		In practice, we often work with a shape-only version of the cross-section, which we can compare to the predictions of various models without the additional uncertainty of the flux prediction influencing the comparison overmuch, as in fig. \ref{fig:1trk ccqe Q2}.  The model most strongly preferred by the data in both $\nu$ and $\bar{\nu}$ cases is not GENIE's simple RFG alone, either with the world average of the free parameter $M_{A}$ (0.99) or with a larger value preferred by fits by MiniBooNE (1.35).  Instead, the data best favor an RFG that is modified using empirical corrections to the cross-section derived from measurements of electron-nucleus scattering (which are thought to be necessitated by inter-nucleon correlations within the nucleus).  The substantial difference from the RFG implemented in the GENIE generator (15-20\% in some regions) underscores the necessity of moving beyond the unadorned RFG model.  Collaborators on the T2K experiment have begun using these data to inform the choice of models used in the NEUT generator to simulate neutrino interactions as well.\cite{CallumT2K}
		
\begin{figure}[h]
			\centering
			\begin{subfigure}[t]{0.49\textwidth}
				\includegraphics[width=\textwidth]{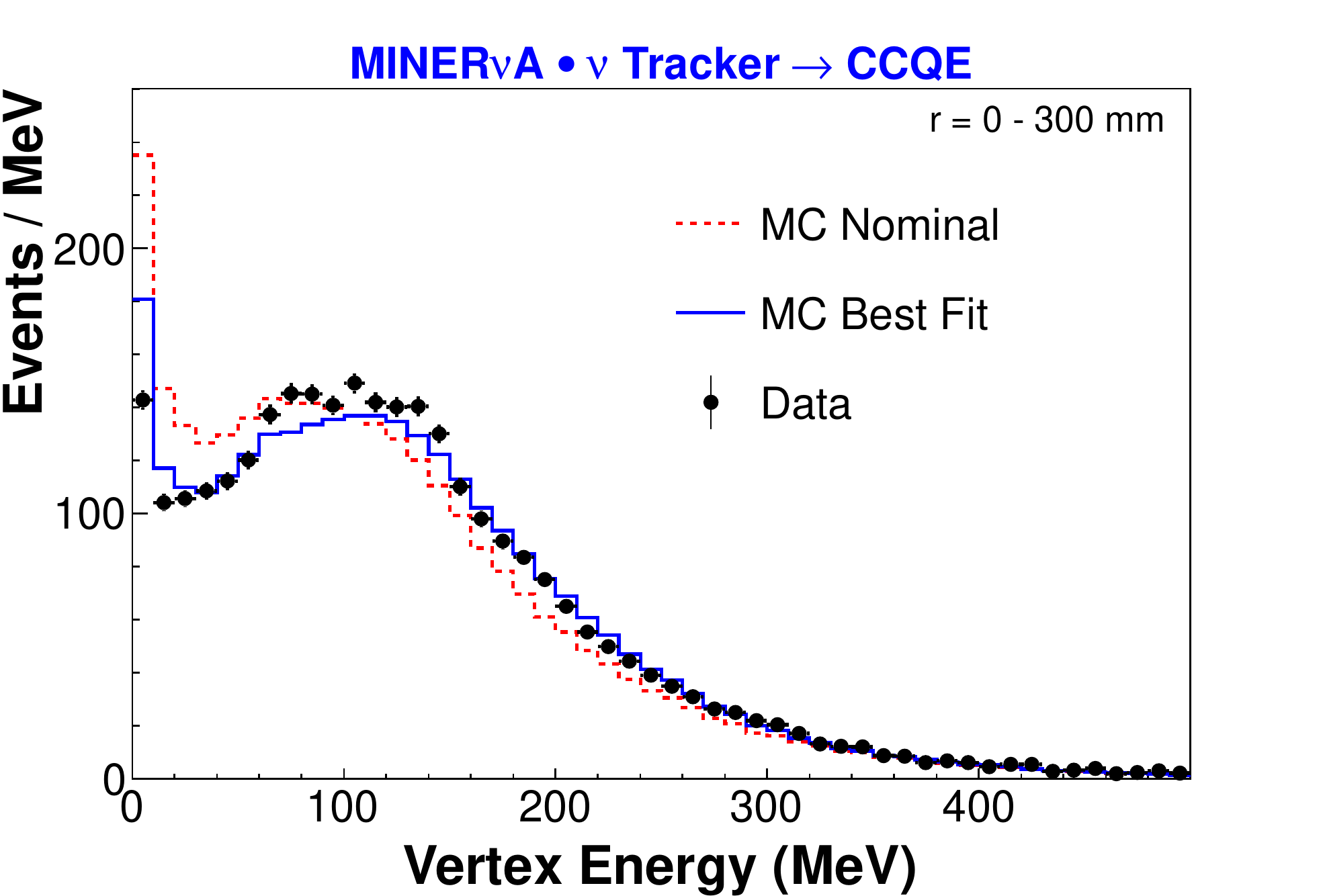}
				\caption{$\nu$ scattering}
				\label{fig:ccqe vertex E nu}
			\end{subfigure}
			\begin{subfigure}[t]{0.49\textwidth}
				\includegraphics[width=\textwidth]{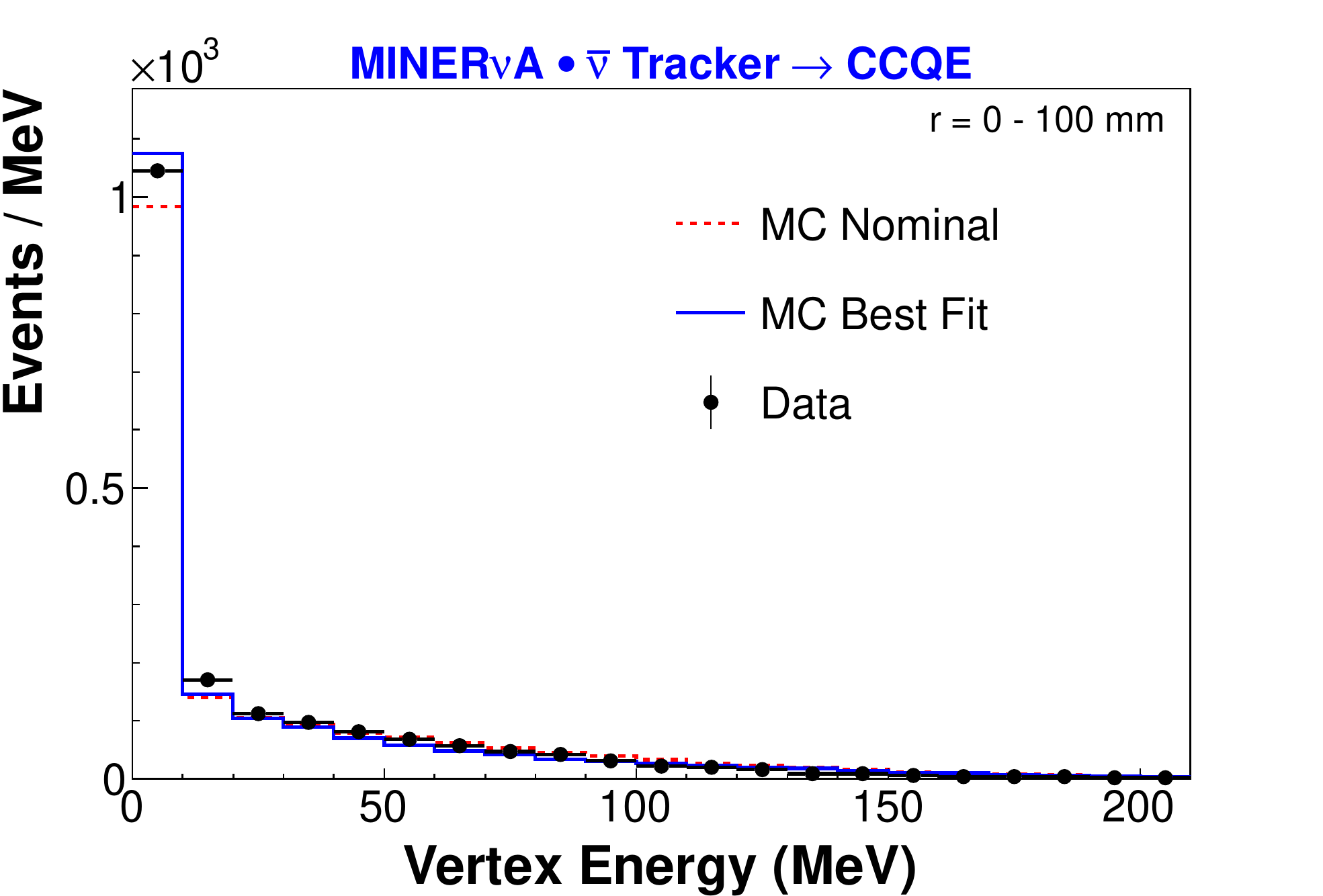}
				\caption{$\bar{\nu}$ scattering}
				\label{fig:ccqe vertex E antinu}
			\end{subfigure}
			\caption{Comparison of GENIE RFG prediction (before and after fitting) of energy near reconstructed CCQE event vertex vs. \mnv{} data.  The fit used for ``Best Fit'' adds an additional proton's energy and is further described in the text.}
		\end{figure}

		To further investigate the hypothesis of correlated nucleon initial states, we also examine the recoil energy spectrum in the vicinity of the reconstructed muon vertex (with displacement $r < 300mm$ for $\nu$ and $r<100mm$ for $\bar{\nu}$) in candidate signal events.  Fig. \ref{fig:ccqe vertex E nu} illustrates the discrepancy between the data distribution and that predicted by the GENIE RFG model for neutrino scattering; it also shows the improvement garnered by adding an additional proton's energy to 25\% of the interactions in the simulated sample.  (We performed a fitting procedure between the data and simulation to determine the fraction of events that should be modified; the result was ($25 \pm 10$)\%.)  We note the constrasting behavior of anti-neutrino scattering, shown in fig. \ref{fig:ccqe vertex E antinu}: here the discrepancy is marginal, and the fitting procedure favored a modification of $(-10 \pm 7)$\% of events, which is consistent with no modification at all.  Taken together, these results lend credence to the picture of initial-state correlations between neutron-proton pairs in some events: neutrino scattering converts a neutron to a proton, resulting in a post-reaction $pp$ pair, and thus an excess proton which we observe in our data but not our simulation; antineutrino scattering, on the other hand, creates a $nn$ pair, and the additional neutron typically (but not always) goes undetected in \mnv{}, so the distributions agree better without modification.
		
\begin{figure}[h]
			\centering
			\includegraphics[width=0.7\textwidth]{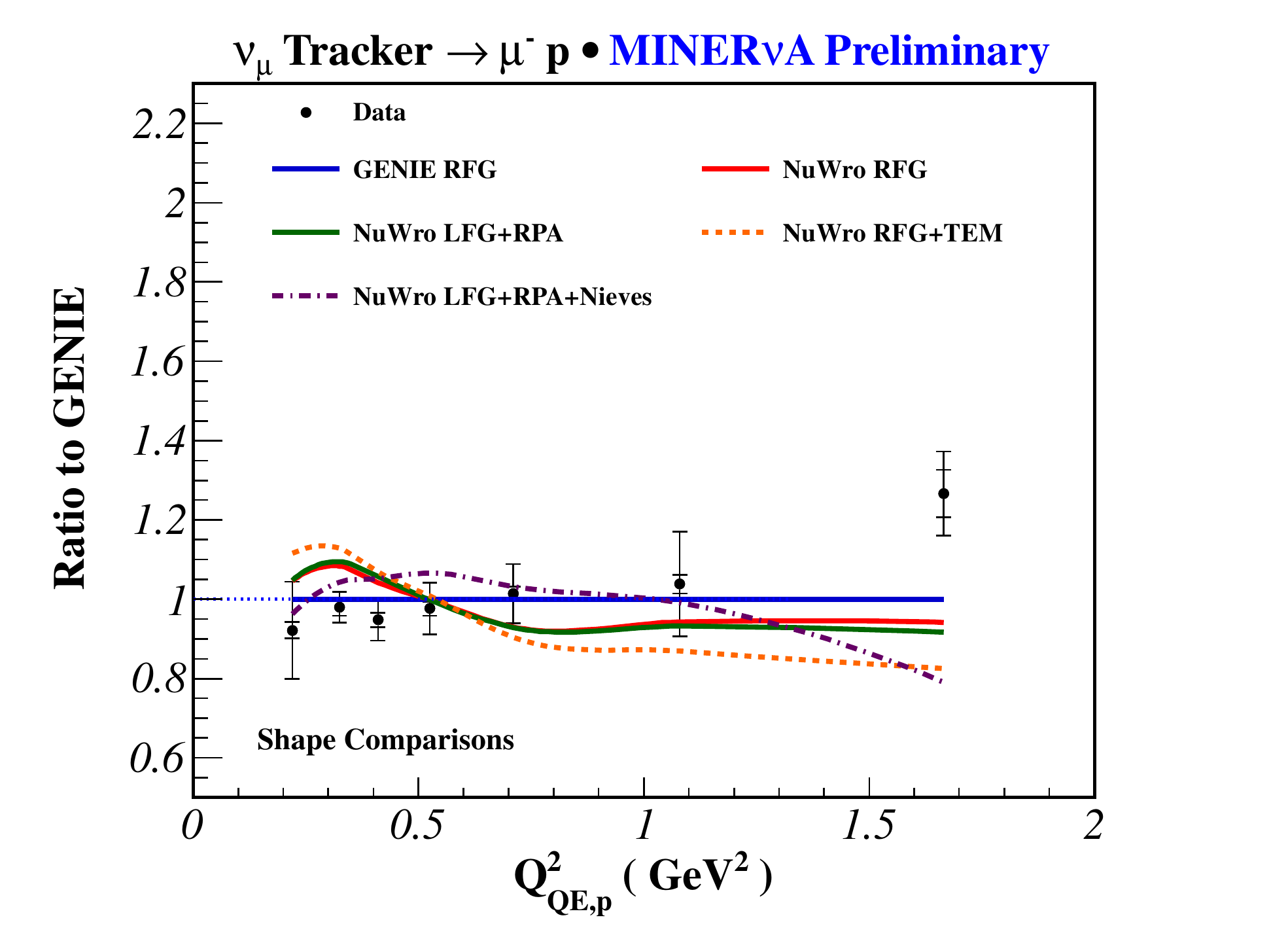}

			\caption{Ratios of the differential cross-section vs. $Q^{2}_{p}$ for \mnv{} data and several models to the usual RFG model as implemented in GENIE.  ``LFG'' means ``local Fermi gas''; ``RPA'' is a random phase approximation\cite{RPA model}; ``Nieves'' is the nucleon correlation model of Nieves et al.\cite{Nieves model}; TEM is a transverse enhancement model\cite{TEM model}.}
			\label{fig:ccqe Q2 proton}
		\end{figure}

		Our second CCQE analysis seeks to approach the kinematic measurement from a different direction.  Most of the selection criteria are analogous to those detailed above.  But instead of demanding that the muon track be analyzed in MINOS, this analysis simply requires the muon candidate to exit the detector (itself a reasonably pure selector for muons).  More significantly, for this measurement we also require that the final-state nucleon's trajectory be reconstructed and that its $dE/dx$ profile positively identify it.  (Since neutrons are typically unobserved in \mnv{}, this therefore restricts the measurement to neutrino mode only, where the final-state nucleon is a proton.)  We then can make the comparison corresponding to fig. \ref{fig:1trk ccqe Q2}, except that we choose to use the \textit{proton}'s kinematics (including its kinetic energy, $T_{p}$) to reconstruct $Q^{2}_{QE}$:
		\begin{equation}
			Q^2_{QE,p} = \left(M_{n} - E_b\right)^{2} + 2\left(M_{n} - E_b\right)(T_{p} - M_{n} - E_{b}) - M_{p}^{2}  
		\end{equation}
		Fig.  \ref{fig:ccqe Q2 proton} shows this comparison.  Here, intriguingly, most of the models fit the data similarly well, with the GENIE RFG model being the slight favorite.  No single explanation for the stark contrast with the result in the muon kinematics (fig. \ref{fig:1trk ccqe Q2}) has yet gained favor, though suggestions include the potential inaccuracy of GENIE's background model as well as a need for better-developed models of multinucleon effects implemented in generators.\cite{2trk paper}
		
		Work on additional cross-section results for the CCQE process is currently underway.  A doubly differential cross-section ($\frac{d^{2}\sigma}{dp_{z}dp_{t}}$) analysis, which will examine the validity of the model predictions in more detail, is in preparation.  A preliminary result for the \textit{electron} neutrino CCQE cross-section (total cross-section vs. energy, as well as differential cross-section vs. $\theta_{e}$, $E_{e}$, and $Q^{2}$) was also presented at this conference.\cite{NuECCQE}
		
	\section{Final-state models: $\pi^{\pm}$ production}
		Since charged pions are both light and strongly interacting, single charged pion production by neutrinos (in \mnv{}, $\nu_{\mu} CH \rightarrow \mu^{-} \pi^{\pm} X$, where $X$ is anything but a $\pi^{\pm}$) is an ideal reaction channel for the study of the final-state (re-)interactions (FSI) of hadrons after a neutrino interaction.  Comparisons of detailed measurements from the MiniBooNE experiment to the predictions of various generator models have raised questions about whether current approaches to FSI modeling are even valid.\cite{GiBUU FSI}  Thus, further measurements are extremely desirable for confirmation or disambiguation of this situation.
		
		For our analysis\cite{chgpi paper}, we select a sample of events containing a charged-current interaction with a single charged pion in the final state by requiring a MINOS-matched muon candidate and one or two other reconstructed tracks.  Exactly one of these tracks must both be consistent with a pion's $dE/dx$ energy loss profile and also be associated with spatially close, time-delayed activity consistent with a decay electron.  We also cut on the reconstructed hadronic invariant mass $W = \sqrt{M_{p}^{2} - Q^{2} + 2M_{p}E_{had}}$ (assuming target nucleon at rest) of the reaction, requiring $W < 1.4$ GeV, so as to select a sample purer in single pion events (typically resulting from decay of the $\Delta$ resonances with $W \sim 1.2$ GeV).
		
\begin{figure}[h]
			\centering
			\includegraphics[width=0.6\textwidth]{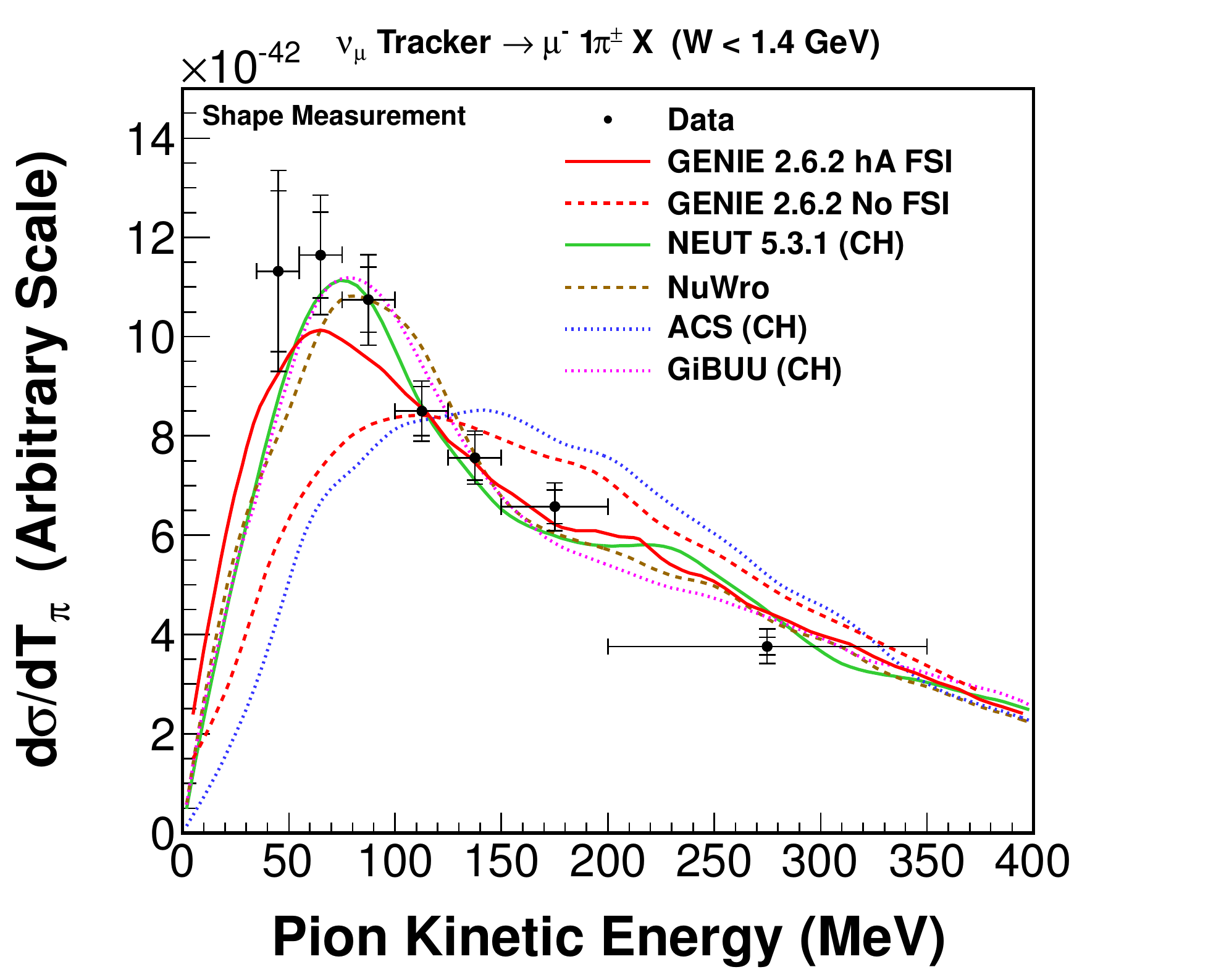}

			\caption{Comparison of various predictions for $d\sigma/dT_{\pi}$ with and without FSI vs. \mnv{} data.  GENIE FSI, NEUT, NuWro, and GiBUU have full FSI models; GENIE No FSI and ACS have no or partial FSI.  (Models are described in more detail in ref. \cite{chgpi paper}.)}
			\label{fig:chgpi}
		\end{figure}

		After extracting shape-only cross-sections vs. pion kinetic energy after the manner of eq. \ref{eq:dsigma}, we are able to compare our data to models which use various different FSI approaches or tunings.  This is shown in fig. \ref{fig:chgpi}.  Models with a full treatment of FSI are strongly preferred by our data, as opposed to models with no or partial FSI simulation.  Once again, collaborators on the T2K experiment are using these data to constrain the FSI parameters used in the NEUT generator to simulate neutrino interactions.

	\section{Cross-section models: coherent $\pi^{\pm}$}
		Finally, we study coherent pion production, in which the momentum transferred from the neutrino is entirely used to create a new particle, and the struck nucleus remains in its ground state.  In its charged form ($\nu_{\mu} A \rightarrow \mu^{-} \pi^{+} A$, or, exchanging leptons for antileptons in antineutrino scattering), this process is a background for $\nu_{\mu}$ CCQE processes; the neutral form ($\nu_{\mu} A \rightarrow \nu_{\mu} \pi^{0} A$) can mimic $\nu_{e}$ CCQE when one of the decay photons from the $\pi^{0}$ is unobserved.
		
\begin{figure}[h]
			\centering
			\includegraphics[width=0.6\textwidth]{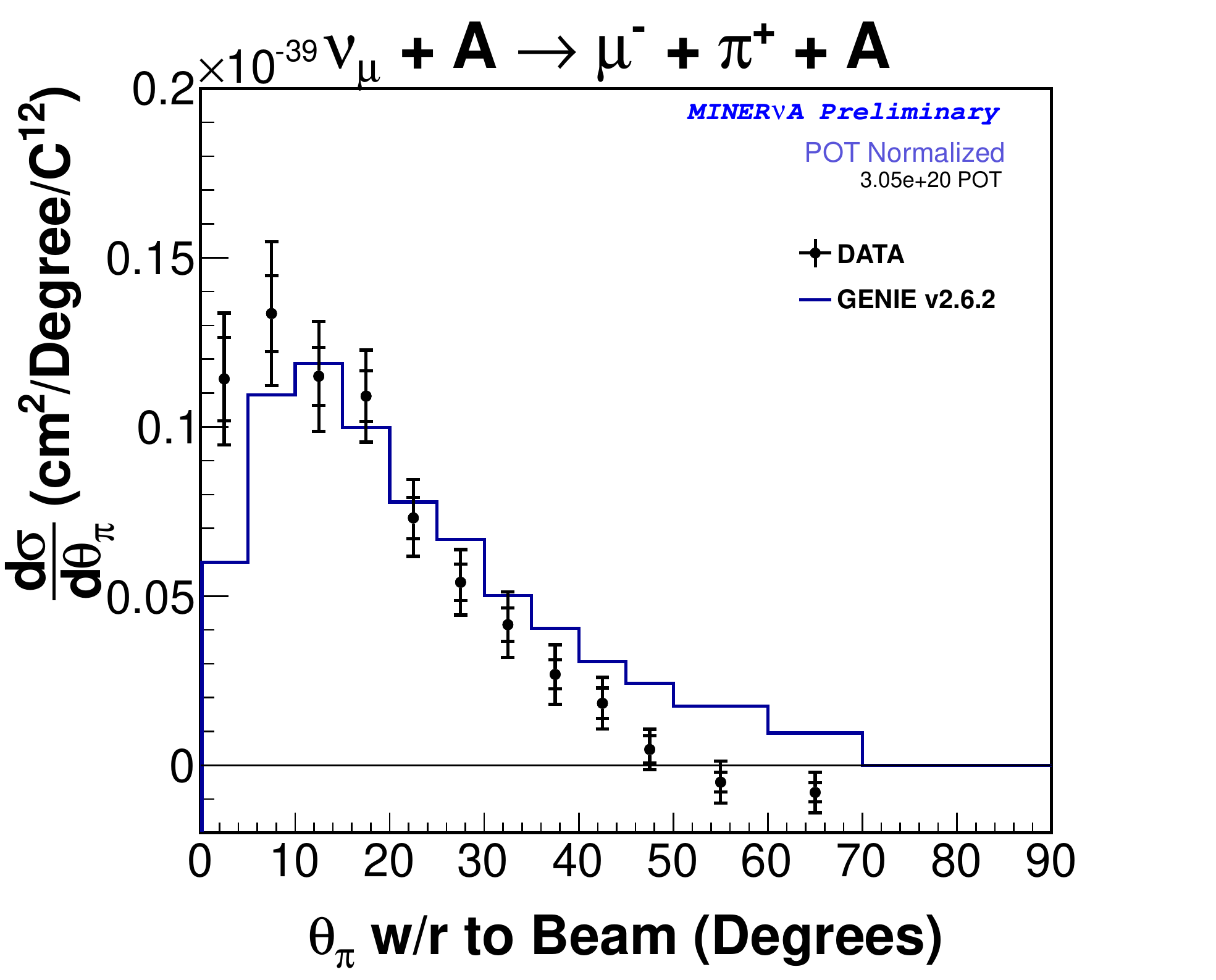}
			\caption{Comparison of the Rein and Sehgal model implemented in GENIE for $d\sigma/d\theta_{\pi}$ vs. \mnv{} data.}
			\label{fig:coh}
		\end{figure}

		For the \mnv{} analysis\cite{coherent paper}, we begin selecting events consistent with charged-current coherent scattering by demanding they have exactly two reconstructed tracks.  We further refine the sample by cutting on the reconstructed momentum transferred to the nucleus, $|t|$, and the energy observed within a ``box'' of side length $40cm$ and depth $15cm$ centered around the reconstructed event vertex.  As before, we construct cross-sections in pion angle according to eq. \ref{eq:dsigma}, and then compare them to the most commonly used model for coherent pion production, that of of Rein and Sehgal, as implemented in GENIE\cite{GENIE}.  This is given in fig. \ref{fig:coh}.  We observe that the measured distributions are significantly more forward than the model in both neutrino (shown) and anti-neutrino scattering (not shown here).  Further discussion of this result and its implications were given in a separate presentation at this conference.\cite{JorgeCoherent}
	
	\section{Conclusions}
		Measurements from \mnv{} provide valuable information in the quest to improve models of neutrino interaction processes and neutrino energy estimation.  We find inadequacies in the simple RFG picture usually used for CCQE, as well as in the underlying cross-section model for coherent pion production, while models of FSI currently used in generators appear to be sufficient to model our data.  These results are already being applied by the T2K collaboration in their generator model tuning, and we anticipate that other generator efforts can benefit from similar studies.  We also expect both forthcoming and planned work (in the low-energy beam configuration used in results discussed here, as well as the medium-energy tune in which \mnv{} is currently collecting data) to continue to improve the fidelity of neutrino interaction modeling, and therefore the quality of reconstructed energy distributions.

\end{document}